\documentstyle[aps,epsfig]{revtex}

\begin{document}
\begin{center}
{\Large \bf
Event-by-Event Fluctuations of Particle Ratios in Central Pb+Pb Collisions at 20 to 158 $A$GeV
}
\end{center}

%\vspace{0.5cm}
\noindent
C.~Roland$^{5}$\footnote{talk presented at Quark Matter 2004}, 
C.~Alt$^{9}$, T.~Anticic$^{21}$, B.~Baatar$^{8}$,D.~Barna$^{4}$,
J.~Bartke$^{6}$, 
L.~Betev$^{9,10}$, H.~Bia{\l}\-kowska$^{19}$, A.~Billmeier$^{9}$,
C.~Blume$^{9}$,  B.~Boimska$^{19}$
% ...
\\
\begin{center}
(NA49 Collaboration)
\end{center}
%\vspace{0.5cm}
\noindent
% ...
$^{4}$KFKI Research Institute for Particle and Nuclear Physics, Budapest, Hungary.\\
$^{5}$MIT, Cambridge, USA.\\
$^{6}$Institute of Nuclear Physics, Cracow, Poland.\\
% ...
$^{8}$Joint Institute for Nuclear Research, Dubna, Russia.\\
$^{9}$Fachbereich Physik der Universit\"{a}t, Frankfurt, Germany.\\
$^{10}$CERN, Geneva, Switzerland.\\
% ...
$^{19}$Institute for Nuclear Studies, Warsaw, Poland.\\
% ...
$^{21}$Rudjer Boskovic Institute, Zagreb, Croatia.\\
\begin{abstract}
In the vicinity of the QCD phase transition, critical fluctuations have been predicted to
lead to non-statistical fluctuations of particle ratios, depending on the nature 
of the phase transition. 
Recent results of the NA49 energy scan program show a sharp maximum
of the ratio of $K^+$ to $\pi^+$ yields in central 
Pb+Pb collisions at beam energies of 20-30 $A$GeV. 
This observation has been interpreted as an indication of a phase transition 
at low SPS energies. We present first results on event-by-event fluctuations 
of the kaon to pion and proton to pion ratios at beam energies close to this maximum.\\
\end{abstract}
Recently the energy scan program at the CERN SPS providing Pb+Pb collisions at beam energies 
of 20, 30, 40, 80 and 158 $A$GeV has been completed. The NA49 collaboration has measured 
identified charged particle spectra at all provided beam energies. One of the most striking 
observations in this data set is a 
sharp maximum of the ratio of average $K^+$ to $\pi^+$ yields in the range of 
20-30 $A$GeV incident beam energy \cite{marekqm04}. Within a statistical model this maximum 
can be interpreted as an indication of the onset of a deconfinement phase transition.
A potential test for the hypothesis of observing direct effects of a phase transition is the 
analysis of event-by-event fluctuations in the hadro-chemical composition of the particle source.
It has been suggested that, depending on the nature and order of the phase transition, 
anomalies in the energy dependence of event-by-event fluctuations might 
be observed and they may allow to characterize the transition process
\cite{marekmult,marekk2pi,strange_hic,stephanov99}.
\begin{figure}[htp]
\centerline{
\epsfig{file=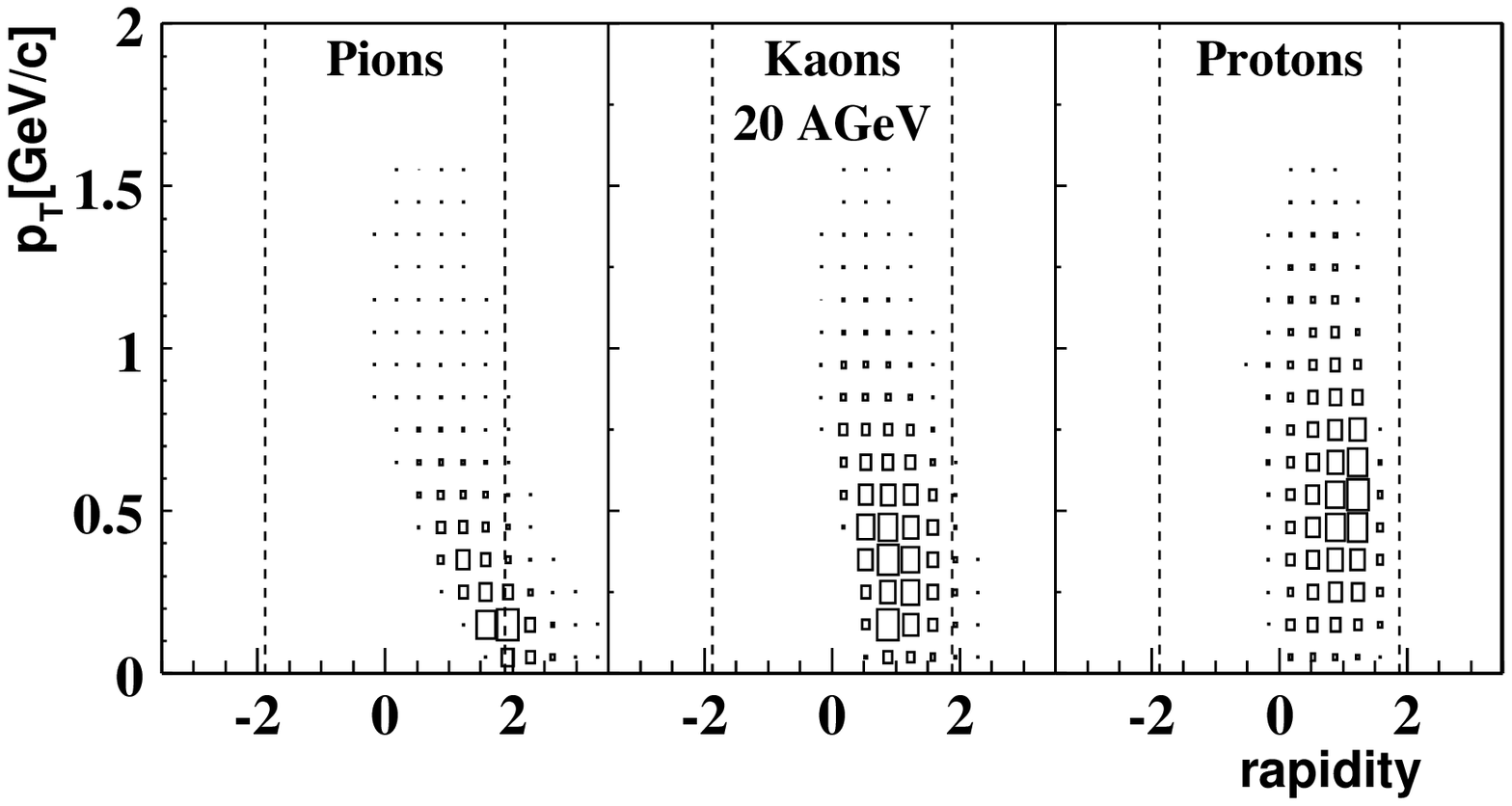,height=4cm,width=12.0cm}
}
\centerline{
\epsfig{file=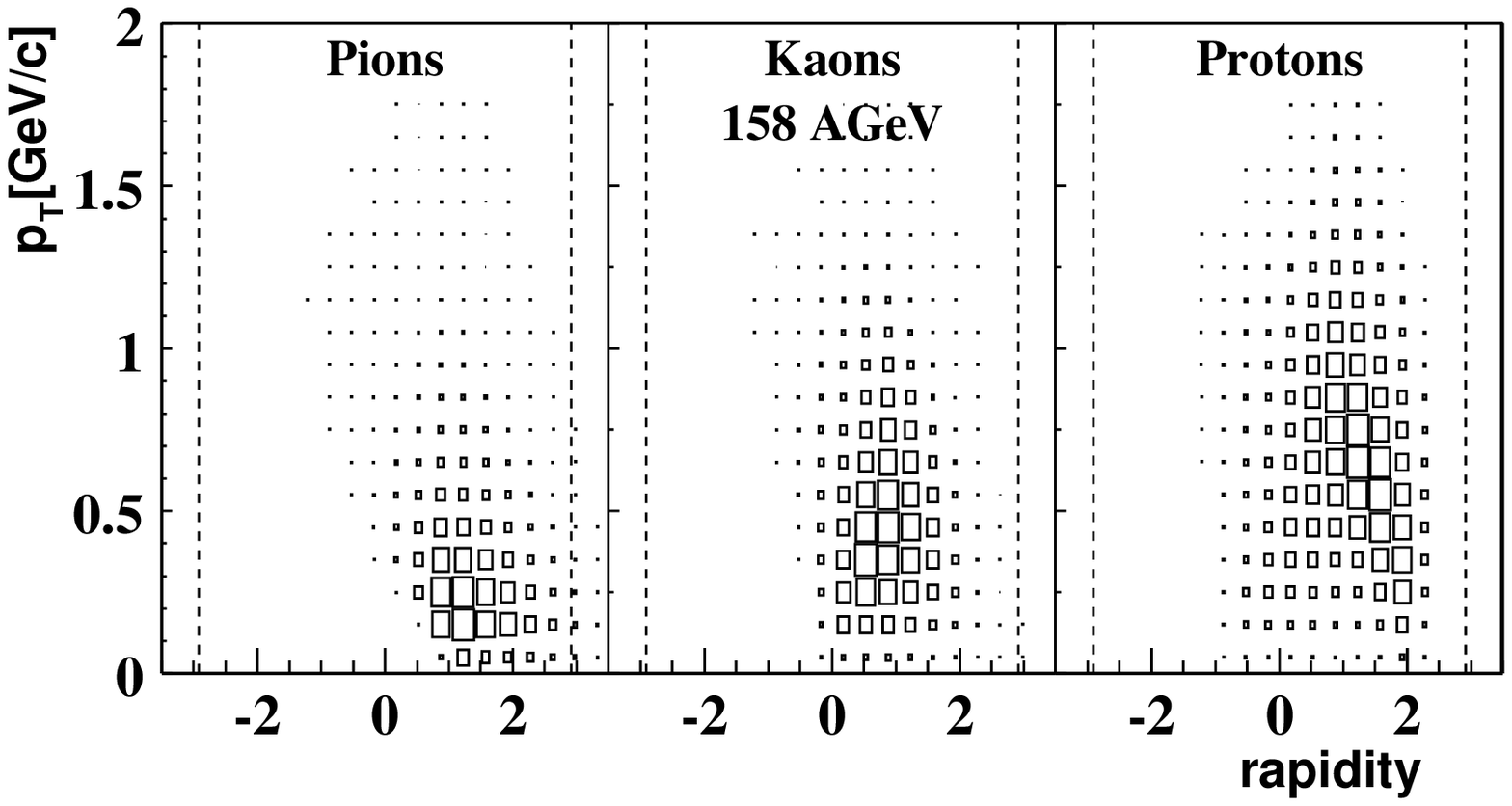,height=4cm,width=12.0cm}
}
\caption{Phase-space distribution of accepted charged particles at 20 $A$GeV (upper panel)
and 158 $A$GeV (lower panel). From left to right pions, kaons and protons are shown. 
No acceptance correction was applied to the particle ratios reported here.}
\label{acc}
\end{figure}
A previous analysis of the event wise ratio of charged kaons to charged
pions ($[K^{+}+K^{-}]/[\pi^{+}+\pi^{-}]$) was performed in Pb+Pb collisions at 158 $A$GeV.
The non-statistical fluctuations of $2.8\%$ observed in the data 
were found to be significantly smaller 
than those expected for an independent superposition of nucleon-nucleon collisions \cite{oldk2pi}.
This supports the interpretation that at the top SPS energy each collision samples the 
same flavor ratios as described in a grand canonical ensemble, combined with a smooth transition from
a possible partonic state to the final state hadronic particle composition. 
The minimal fluctuations expected due to production of the final 
state hadrons via resonances completely exhaust the observed fluctuation signal \cite{koch99}.

In this article the extension of this analysis to lower incident energies is presented.
In addition to the estimate of the $[K^{+}+K^{-}]/[\pi^{+}+\pi^{-}]$ ratio also the ratio 
of $[p+\bar{p}]/[\pi^{+}+\pi^{-}]$ is determined on an event-by-event basis. 
The analysis method is described in \cite{oldk2pi,rolandc99,gazdzicki95}. 
The extension of this method to fitting multiple particle 
ratios for each event and the robustness of the event-by-event ratio estimators against effects 
of very low particle multiplicities have been validated by extensive systematic checks 
and Monte Carlo studies.

\begin{figure}[htp]
\centerline{
\epsfig{file=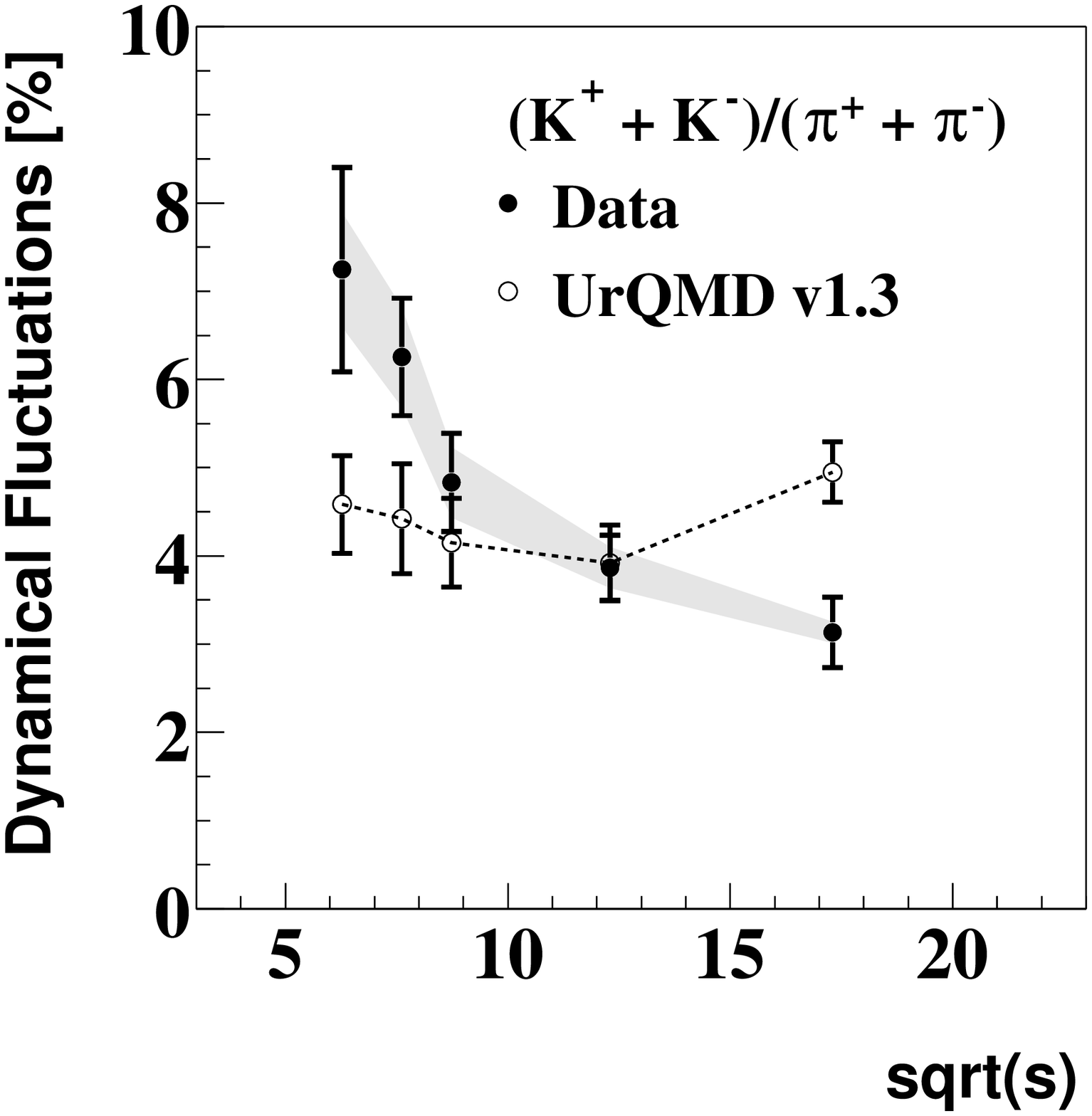,height=5.0cm}
\epsfig{file=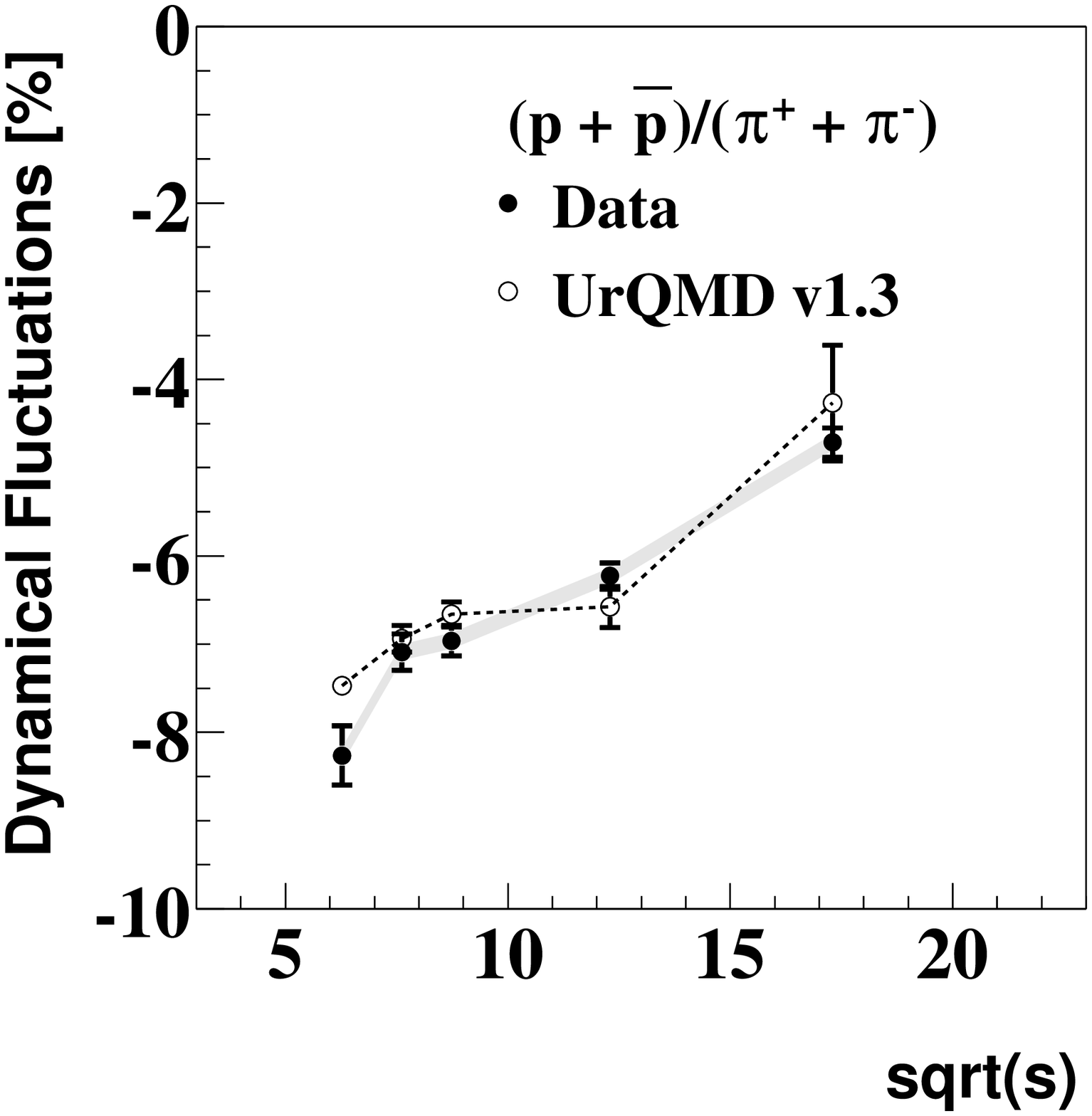,height=5.0cm}
}
\caption{ Energy dependence of the event-by-event fluctuation signal of the $[K^{+}+K^{-}]/[\pi^{+}+\pi^{-}]$ ratio (left panel) and the $[p+\bar{p}]/[\pi^{+}+\pi^{-}]$ ratio (right panel). 
The systematic errors of the measurements are shown as gray bands.}
\label{edep}
\end{figure}

A detailed description of the NA49 experiment can be found in \cite{nimpaper}.
At the five available beam energies the 3.5\% most central Pb+Pb collisions
were selected based on projectile spectator energy. 
Figure \ref{acc} shows the phase space distribution of accepted pions, kaons and protons 
at 158 $A$GeV compared to the 20 $A$GeV data set. The shifting cut off for pions is due to the 
requirement of a minimum total momentum of 3 GeV/c in order to facilitate 
the particle identification.
The relative width $\sigma$, defined as $\sigma = RMS/Mean*100~[\%]$, of the 
measured event-by-event particle ratio distributions can be decomposed into 
three contributions:
\begin{enumerate}
\item Due to the finite number of particles produced and observed per event,
the ratio of particle multiplicities measured event-by-event will exhibit
statistical fluctuations with a width dictated by the individual particle
multiplicities.
\item Due to non-ideal particle identification these statistical fluctuations
will be smeared by the experimental $dE/dx$ resolution and the event-by-event
fitting procedure.
\item Superimposed on the background of statistical and experimental fluctuations we may observe
genuine non-statistical fluctuations.
\end{enumerate}
An accurate estimate of the contribution due to finite number fluctuations in the particle multiplicities and effects of detector resolution is obtained using a mixed event technique.
Mixed events are constructed by combining particles randomly selected from different
events, reproducing the multiplicity distribution of the real events. 
\begin{figure}[htp]
\centerline{
\epsfig{file=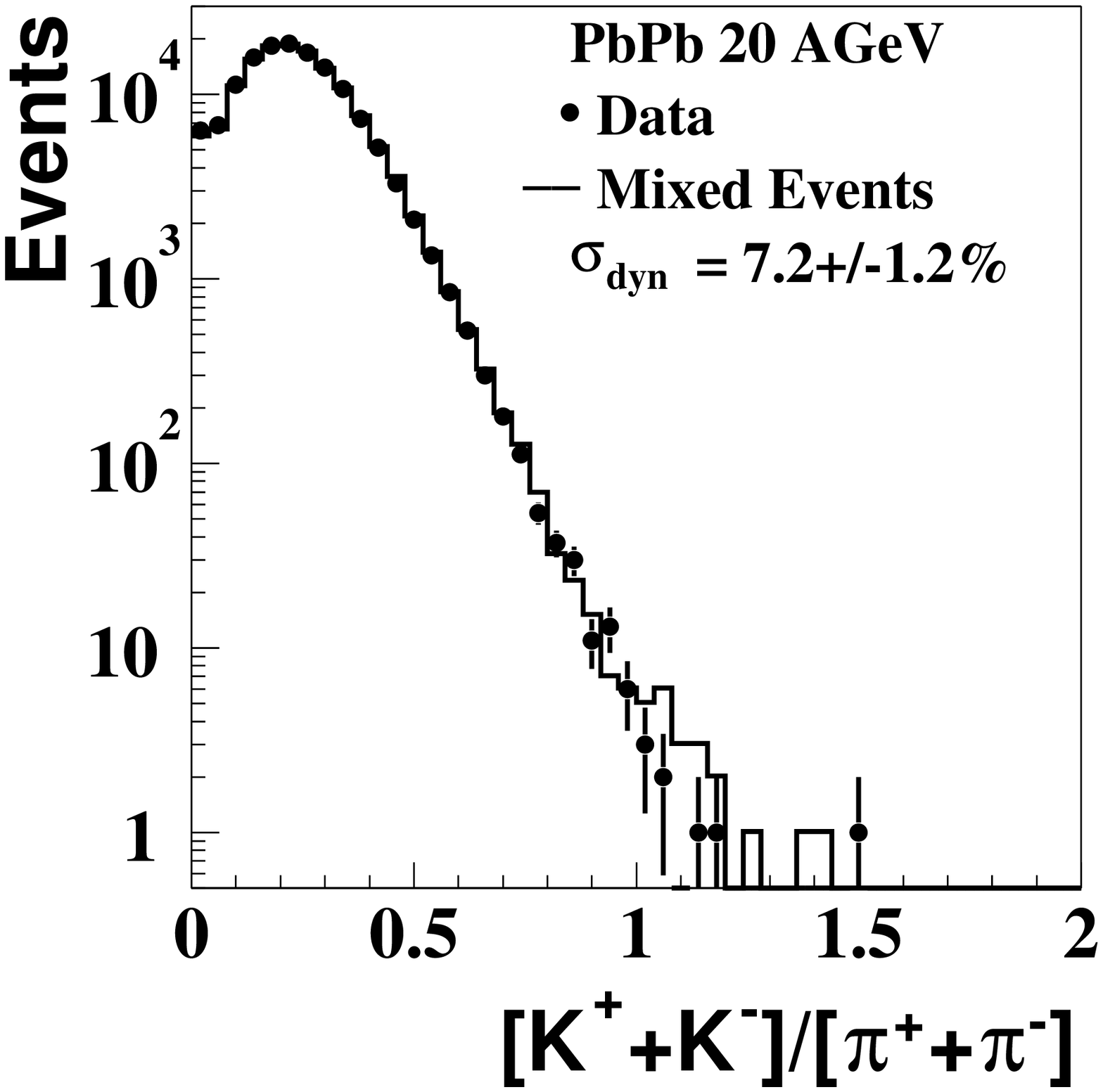,height=3.5cm}
\epsfig{file=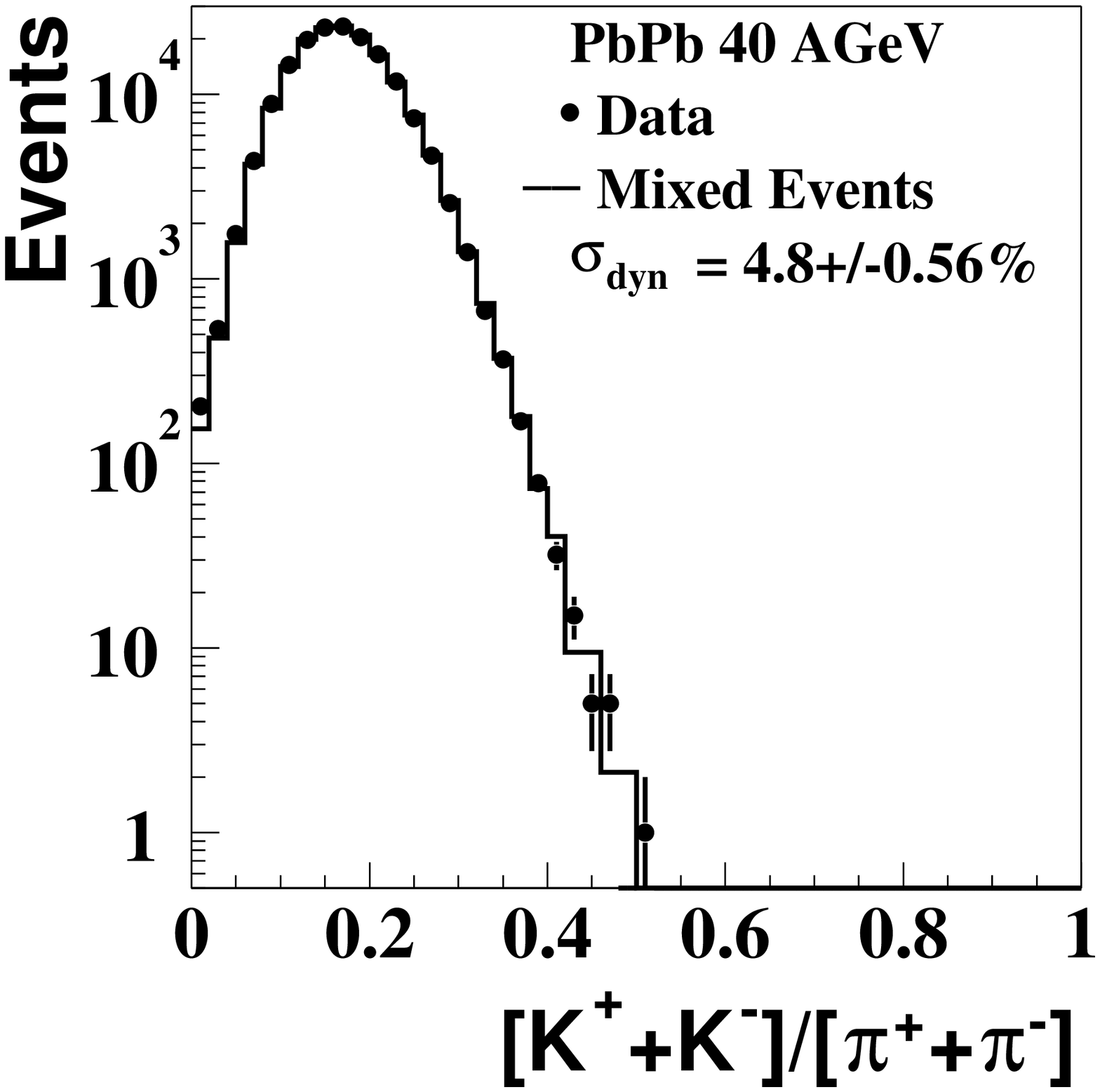,height=3.5cm}
\epsfig{file=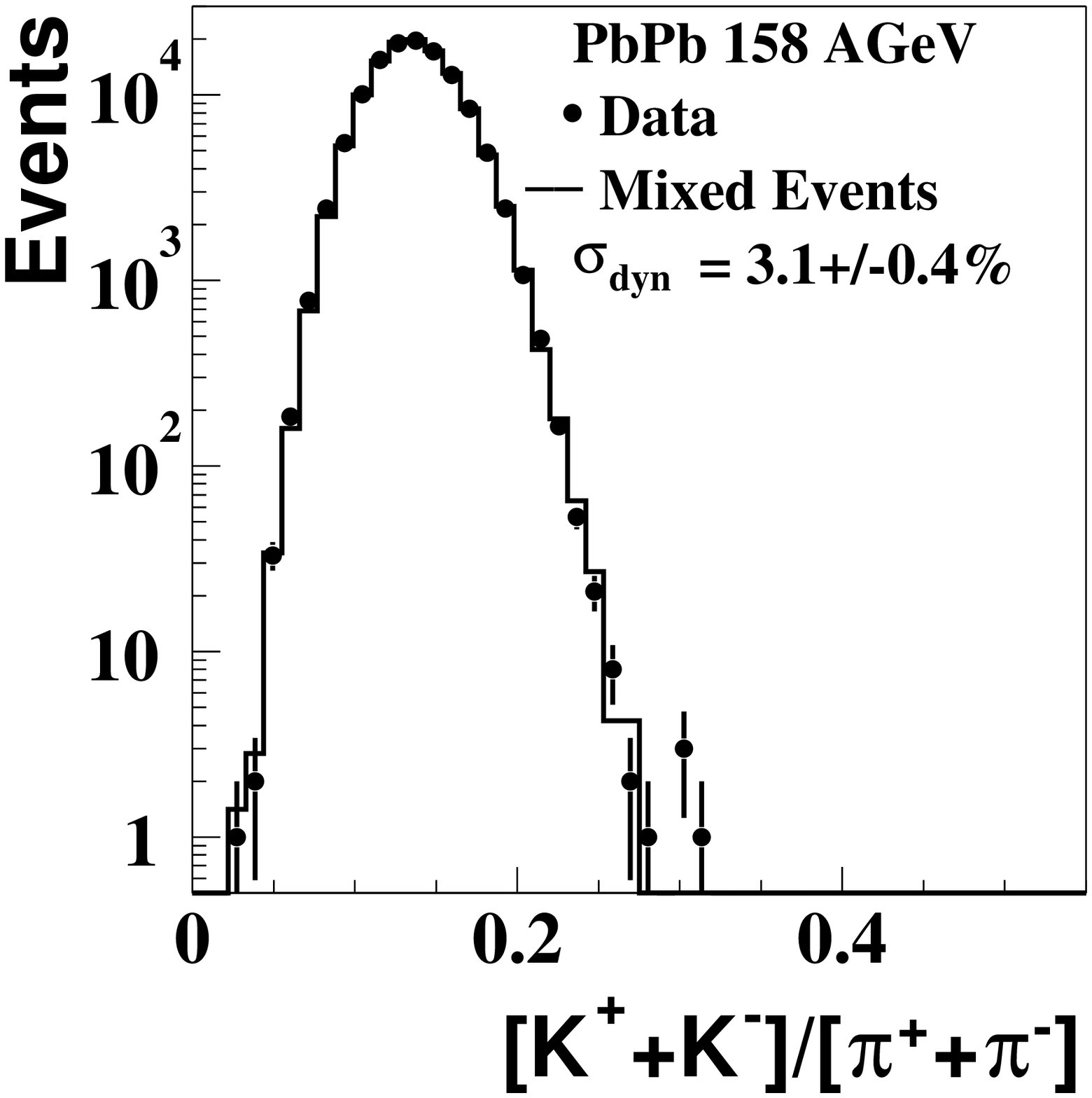,height=3.5cm}
}
\caption{Distributions of the event-by-event $[K^{+}+K^{-}]/[\pi^{+}+\pi^{-}]$ ratio for data (points) and mixed events (histogram).}
\label{k2pi}
\end{figure}
\begin{figure}[htp]
\centerline{
\epsfig{file=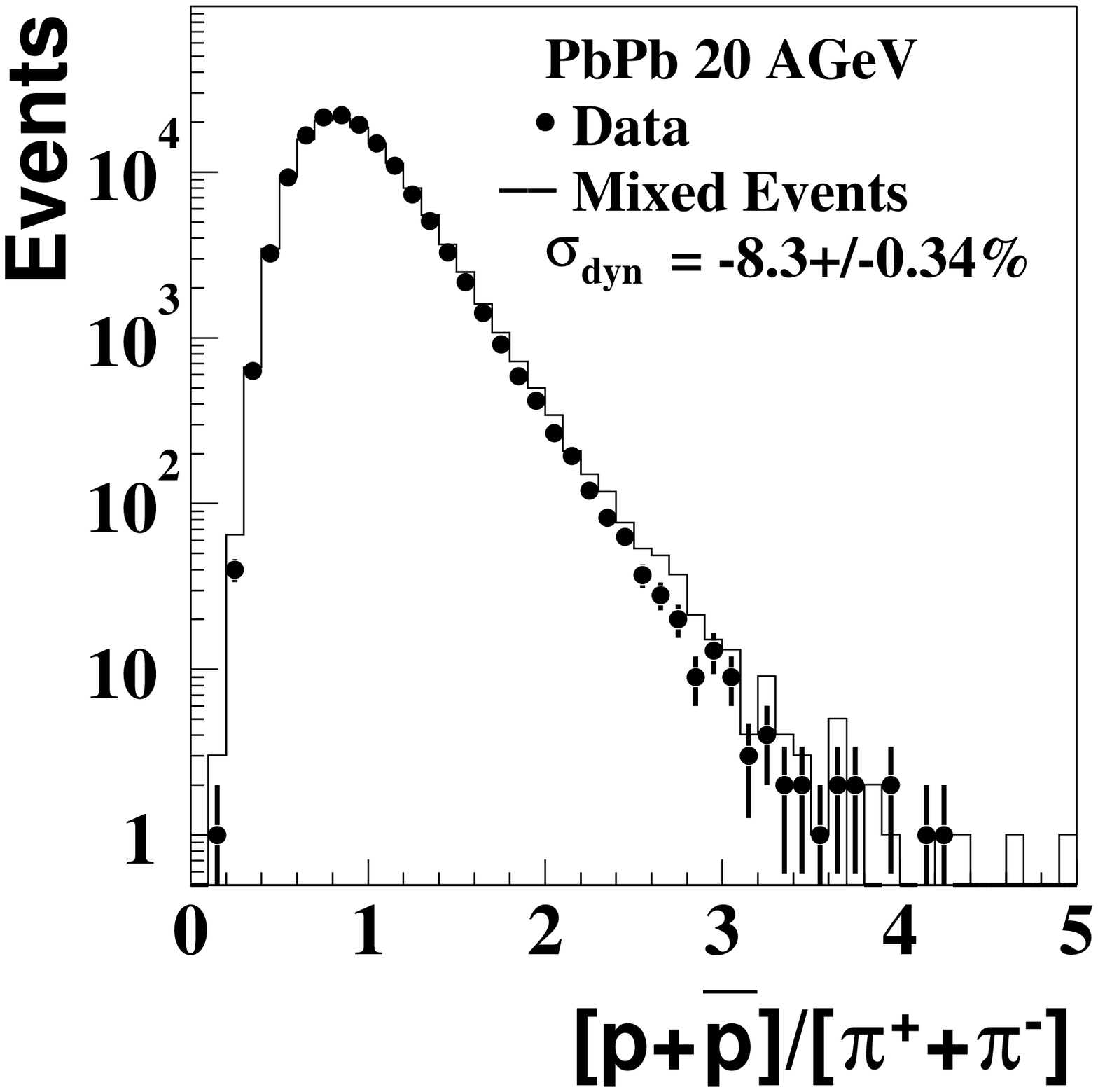,height=3.5cm}
\epsfig{file=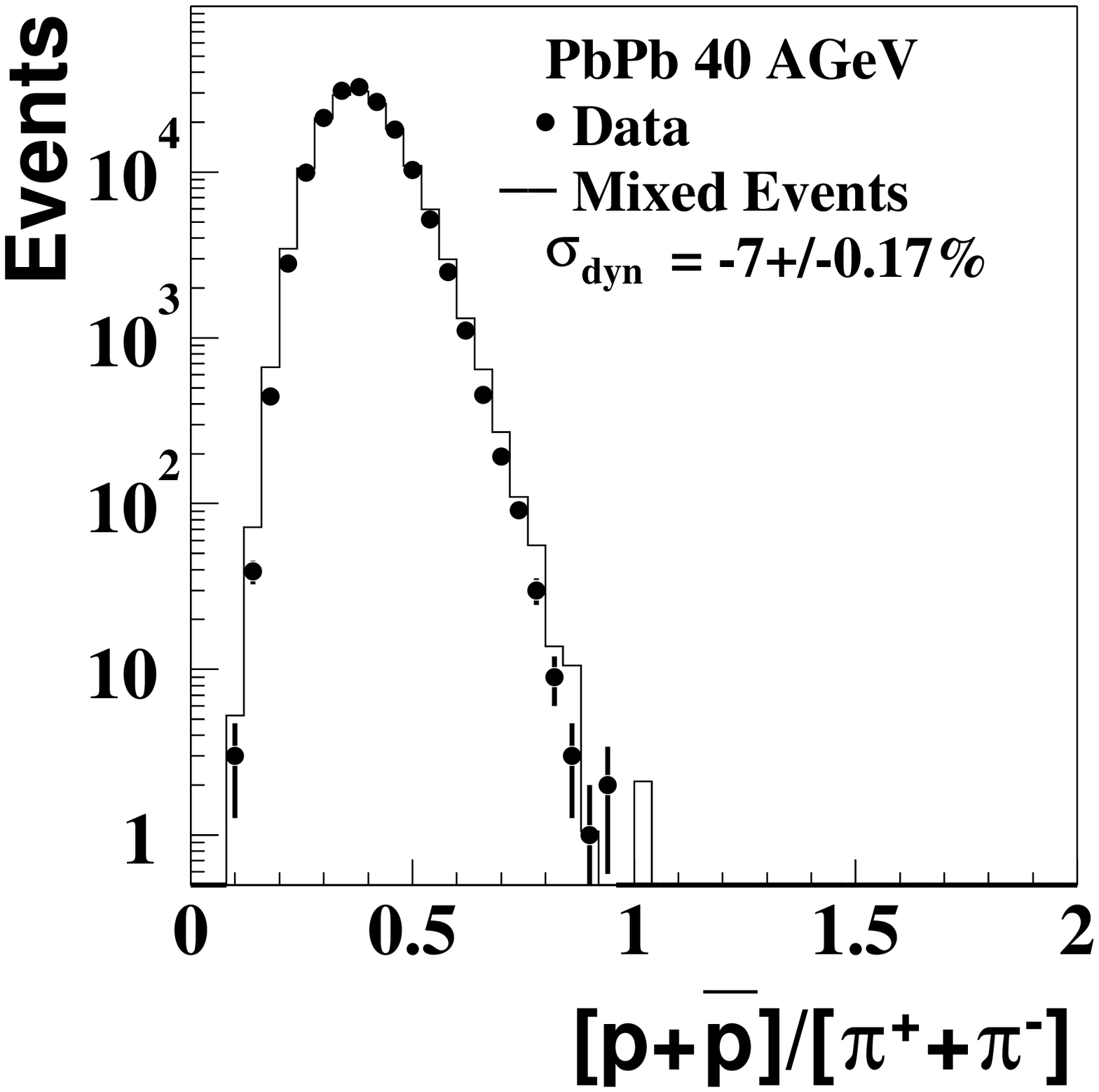,height=3.5cm}
\epsfig{file=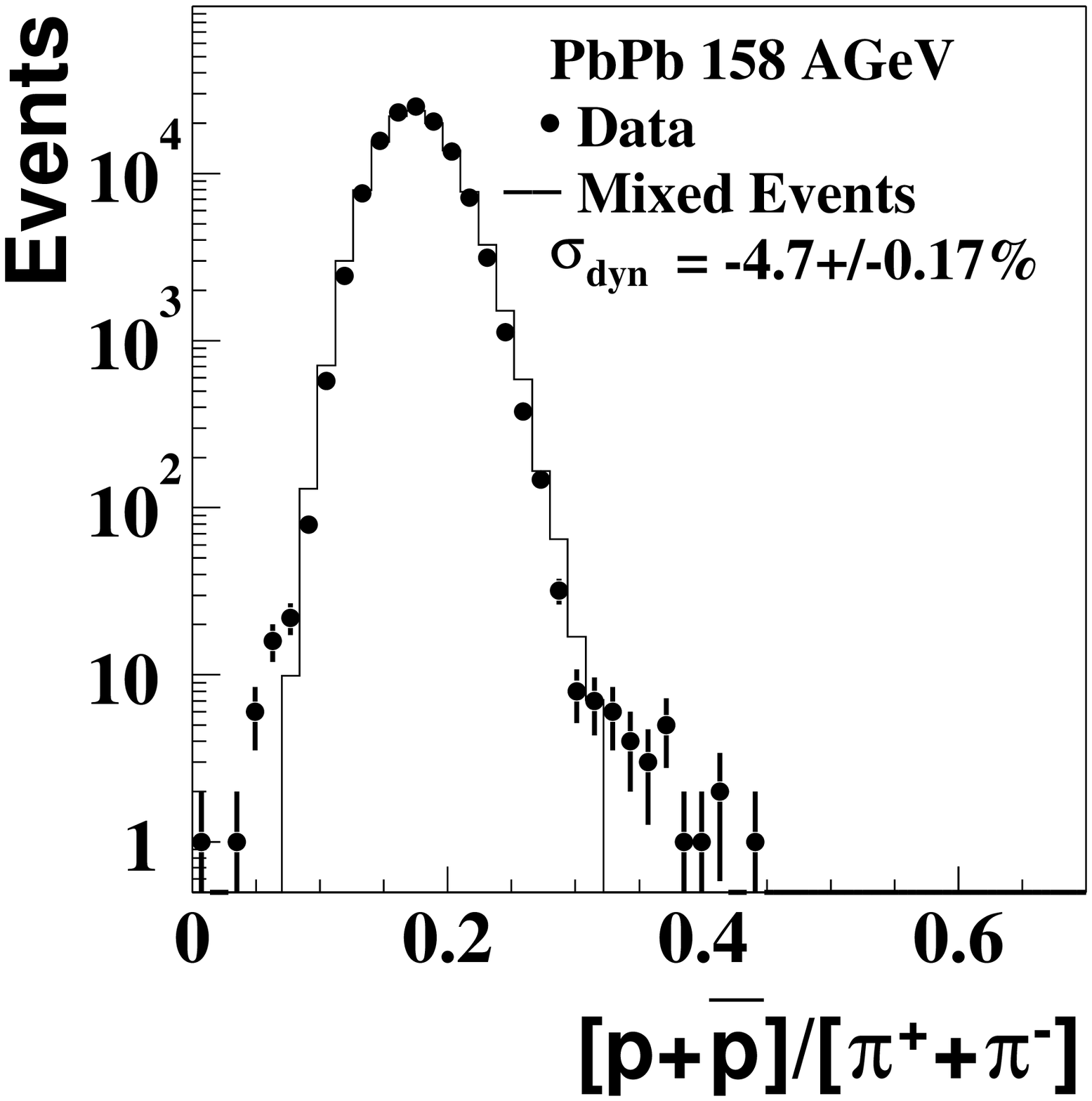,height=3.5cm}
}
\caption{Distributions of the event-by-event  $[p+\bar{p}]/[\pi^{+}+\pi^{-}]$ ratio for data (points) and mixed events (histogram).}
\label{p2pi}
\end{figure}
These mixed events are then subjected to the same fit procedure as the real events.
By construction, the mixed events on average have the same particle
ratios as the real events, but no internal correlations. 
To quantify the relative contribution of dynamical fluctuations $\sigma_{dyn}$, the r.m.s width $\sigma_{mix}$ of the mixed event distribution is subtracted from the r.m.s width $\sigma_{data}$ of 
the real event particle ratio distribution:
\begin{equation}
\sigma_{dyn} = sign(\sigma_{data}^2 - \sigma_{mixed}^2) \sqrt{|\sigma_{data}^2 - \sigma_{mixed}^2|}
\label{sdyn}
\end{equation}
The systematic error of the fluctuation measurement is estimated by comparing results when using two
different track quality cuts. The values of the fluctuation signal presented here are calculated as 
the arithmetic mean of both results.
For illustration the distributions of the two studied particle ratios at 20, 40 and 158 $A$GeV
 are shown in Figs. \ref{k2pi} and \ref{p2pi}. 
The contribution from dynamical fluctuations obtained from Eq. \ref{sdyn} 
is plotted in the Fig. \ref{edep}.
Fluctuations of the $K/\pi$ ratio are positive and decrease with beam energy.
In case of the $[p+\bar{p}]/\pi$ ratio the width of the data distribution is smaller than the width of the distribution of mixed events. The dynamical fluctuations are negative.
A negative fluctuation signal of the event wise $[p+\bar{p}]/\pi$ ratio can be understood, if resonance decays into pions and protons are considered.
The magnitude of the negative fluctuation signal in the $[p+\bar{p}]/\pi$ channel may be 
related to the relative contribution of resonance decay products in the final state of 
the collision.
In order to estimate the significance of the observed fluctuation signals of the two ratios considered, 
we compare the data to a string-hadronic cascade model UrQMD \cite{urqmd}. In this model, by construction, no fluctuations due to a
potential phase transition are present, while resonance decays are included as well as effects 
of correlated particle production due to quantum number and energy/momentum conservation laws. 
For this study, large samples of UrQMD events were generated at all five beam energies and then subjected 
to an acceptance filter modeling the NA49 detector system. 
The accepted final state particles were counted and the corresponding ratios were formed.
The energy dependence of the event-by-event $[p+\bar{p}]/\pi$ ratio in UrQMD closely matches the 
energy dependence observed in the data, as shown in Fig. \ref{edep}. This lends further support to 
interpreting the negative fluctuation signal as resulting from resonance decays.
In case of the fluctuations of the event wise $K/\pi$ ratio, the energy dependence of the signal cannot 
be reproduced in the cascade model. UrQMD gives an energy independent fluctuation signal. Since the relative contribution of resonances changes dramatically with incident beam energy, we conclude that in the 
$K/\pi$ ratio resonances do not give a significant contribution to the fluctuation signal.
The finite fluctuation signal in the UrQMD model can be attributed to correlated particle production 
due to conservation laws.
In the data we observe a significantly smaller fluctuation signal at highest beam energies than 
in a cascade model. At 158 $A$GeV the fluctuation signal is consistent with calculations performed assuming 
a grand canonical ensemble without enforcing local conservation laws. Towards lower beam energies 
we see a steep increase of the fluctuation signal as shown in the left panel of Fig. \ref{edep}.
The increase of the signal goes significantly beyond the value seen in a hadronic cascade model, 
indicating the onset of a new source of fluctuations. 
Further theoretical calculations will be needed to evaluate the relevance of this observation 
for a possible interpretation in the context of a deconfinement phase transition.\\
Acknowledgements: This work was supported by the US Department of Energy
Grant DE-FG03-97ER41020/A000,
the Bundesministerium fur Bildung und Forschung, Germany, 
the Polish State Committee for Scientific Research (2 P03B 130 23, SPB/CERN/P-03/Dz 446/2002-2004, 2 P03B 04123), 
the Hungarian Scientific Research Foundation (T032648, T032293, T043514),
the Hungarian National Science Foundation, OTKA, (F034707),
the Polish-German Foundation, and the Korea Research Foundation Grant (KRF-2003-070-C00015).\\

\end{document}